\documentclass[11pt]{article}
\usepackage{amsmath,amsfonts,latexsym,graphicx,amssymb}
\usepackage[dvips]{epsfig}
\usepackage{amsfonts}
\usepackage{bm}
\date{}

\newcommand{\ot}{{\,\otimes\,}}
\newcommand{{\Cd}}{{\mathbb{C}^d}}

\def\<{\langle}
\def\>{\rangle}

\begin{document}

\title{\bf Parameterizing density matrices for composite quantum systems}
\author{Erwin Br\"uning \\
School of Mathematical Sciences,
University of KwaZulu-Natal, \\
 Westville Campus, Private Bag X54001, Durban 4000, South Africa \\
 \\
Dariusz Chru\'sci\'nski \\
Institute of Physics, Nicolaus Copernicus University, \\
ul. Grudzi\c{a}dzka 5/7,  87-100 Toru\'n, Poland\\ \\
Francesco Petruccione \\
School of Physics, University of KwaZulu-Natal,\\
Westville Campus, Private Bag X54001, Durban 4000, South Africa}

\maketitle

\begin{abstract}

A parametrization of density operators for bipartite quantum
systems is proposed. It is based on the particular parametrization
of the unitary group found recently by Jarlskog.  It is expected
that this parametrization will find interesting applications in
the study of quantum properties of many partite systems.

\end{abstract}


\section{Introduction}

 Density operators represent states of quantum
systems. They are crucial to describe the dynamics of open quantum
systems \cite{BP,Alicki}. Recently, there is a considerable
interest in the structure of the set of density operators due to
the emerging field of quantum information theory \cite{QIT}. It
turns out that quantum entanglement may be used as basic resource
in quantum information processing and communication. The prominent
examples are quantum cryptography, quantum teleportation, quantum
error correction codes and quantum computation. It is, therefore,
clear that the proper description of the set of density operators
is highly important.

It turns out that the structure of the set of density operators is
nontrivial and it is well understood only for 2-level systems
\cite{Nasza,Karol}. This problem was studied by many authors from
different perspectives and various parameterizations of the set of
density matrices were proposed \cite{Hioe}--\cite{Dita0}, see also
the recent paper \cite{BK} for the generalized Bloch vector
approach. Recently, two of us proposed another parametrization
\cite{Erwin} which is based on the particular parametrization of
the unitary group found by Jarlskog \cite{J}. This parametrization
was already applied for modeling quantum gates in the context of
quantum computing \cite{Fujii} and to investigate possible time evolutions for density
matrices \cite{Erwin}.

In the present paper we propose a similar parametrization for
composite systems. This problem is of significant importance for
analyzing quantum entanglement. Although the analysis of states is
parametrization invariant (as much of the problems in physics) a
clever parametrization may considerably simplify the problem and
sheds new light into the structure of quantum states. Moreover, as
usual, the particular parametrization depends very much upon the
problem one would like to analyze.

Recently, a simple parametrization enabling one to analyze the PPT
property (positive partial transpose) was proposed in \cite{SPPT}.
This special parametrization was used to define a new class of
states (so called SPPT states) which are PPT but of a very special
form. It was conjectured that SPPT states are separable. Now, we
study the problem of  parametrization of composite systems from
a more general perspective. It is expected that the proposed
parametrization will be helpful in analyzing the intricate
structure of quantum entanglement.

The paper is organized as follows: in Section \ref{par} we recall
the basic ingredients of the parametrization of $n$-level systems used in
\cite{Erwin}. Then in Section \ref{com} it is shown how to
generalize it for arbitrary $n \ot m$ composed systems. The
procedure is illustrated by explicit examples  where it is easy to
check for separability. Finally, we end with some conclusions.


\section{Parametrization for $n$-level system}
\label{par}

  The parametrization
used in \cite{Erwin} is defined as follows: any density matrix
$\rho_n$ may be written as
\begin{equation}\label{I}
    \rho_n = U_n D(\lambda_1,\ldots,\lambda_n) U_n^\dagger\ ,
\end{equation}
where the matrix $D(\lambda_1,\ldots,\lambda_n)$ is diagonal with
$\lambda_k$ on the main diagonal and $U_n$ is a unitary matrix from
$SU(n)$. Now, following \cite{J} any element $U_n$ from $SU(n)$ may
be factorized as follows
\begin{equation}\label{}
U_n = A^n_{n} A^{n-1}_{n} \ldots A^{2}_{n} A^1_n \ ,
\end{equation}
where
\begin{equation}\label{A-X}
A^{j}_{n}  = e^{X_j}\ ,
\end{equation}
with $X_j \in su(n)$ for $j=1,2,\ldots,n$. The antihermitian
traceless matrices $X_j$ entering (\ref{A-X}) are defined as
follows: $X_1$ is diagonal and the $X_j$, for $j=2,\ldots,n$, are
given by
\begin{equation}\label{Xj}
X_j = \left( \begin{array}{c|c|c} \mathbb{O}_{j-1} & |z_j\> & 0  \\
\hline - \< z_j| & 0 & 0 \\ \hline 0 & 0 & \mathbb{O}_{n-j}\ ,
\end{array} \right) \,
\end{equation}
where $\mathbb{O}_k$ denotes $k \times k$ null matrix, and $|z_j\>$
denotes a complex vector from $\mathbb{C}^{j-1}$
\begin{equation}\label{}
    |z_j\> = \left( \begin{array}{c} z_{1j} \\ \vdots \\ z_{j-1,j}
    \end{array} \right)\ ,
\end{equation}
together with $\<z_j| =
(\overline{z}_{1,j},\ldots,\overline{z}_{j-1,j})$. Taking into
account that $A^1_n$ is diagonal, formula (\ref{I}) implies
\begin{equation}\label{}
    \rho_n = A^n_{n}  \ldots A^{2}_{n}\, D(\lambda_1,\ldots,\lambda_n)\,
    {A^{2}_{n}}^\dagger \ldots  {A^n_{n}}^\dagger\
    .
\end{equation}
Hence, to parameterize $\rho_n$ one needs $(n-1)$ complex vectors
$z_2,\ldots,z_n$, with $z_j \in \mathbb{C}^{j-1}$, and $(n-1)$
real parameters $\lambda_1,\ldots,\lambda_{n-1}$. All together
$(n^2-1)$ real parameters (to parameterize $n \times n$ Hermitian
matrix one needs indeed $n^2$ real parameters -- $n$ on the
diagonal and $n(n-1)$ off-diagonal -- and the normalization
eliminates one parameter).  Now, a simple calculation \cite{Fujii}
gives
\begin{equation}\label{A..}
    A^j_{n} = \left( \begin{array}{cc} V^j_{n}  & 0 \\ 0 & \mathbb{I}_{n-j}
    \end{array} \right) \ ,
\end{equation}
with $\mathbb{I}_k$ being the $k \times k$ unit matrix and $V^j_{n}$
being the $j\times j$ unitary matrix given by
\begin{equation}\label{V..}
    V^j_{n} = \left( \begin{array}{cc} \mathbb{I}_{j-1} -(1-c_j) |\widetilde{z}_j\>\<\widetilde{z}_j|   & s_j |\widetilde{z}_j\> \\
    -s_j \<\widetilde{z}_j| & c_j
    \end{array} \right) \ ,
\end{equation}
where  $|\widetilde{z}_j\>$ denotes the unit vector
\begin{equation}\label{}
    |\widetilde{z}_j\> = \frac{|z_j\>}{||z_j||}\ ,
\end{equation}
that is,
\begin{equation}\label{zz}
    \<\widetilde{z}_j|\widetilde{z}_j\> = ||\widetilde{z}_j||^2=1\ ,
\end{equation}
 and
\begin{equation}\label{}
    c_j := \cos\theta_j \ , \ \ \ \ s_j := \sin\theta_j \ ,
\end{equation}
with
\begin{equation}\label{}
    \theta_j := ||z_j||\ .
\end{equation}
As usual, in Eq. (\ref{V..})  $|\widetilde{z}_j\>\<\widetilde{z}_j|$
denotes the $(j-1)\times (j-1)$ matrix defined by
\begin{equation}\label{}
    (|\widetilde{z}_j\>\<\widetilde{z}_j|)_{kl} :=
    \overline{\widetilde{z}}_k \widetilde{z}_l \ ,
\end{equation}
for $k,l=1,\ldots,j-1$.

 Therefore,  $\rho_n$ is parameterized by $(n-1)$ eigenvalues
$\lambda_1 \geq \ldots \geq \lambda_{n-1}$, $(n-1)$ unit vectors
$\widetilde{z}_2,\ldots,\widetilde{z}_n$, i.e., $\widetilde{z}_j$
defines a point on the  $(2j-1)$--dimensional unit sphere $S^{2j-1}$,
and $(n-1)$ angles $\theta_2,\ldots,\theta_n$ from the hyperoctant
of $(n-1)$--dimensional space. All together we have the correct number of independent parameters of a $n
\times n $ density matrix
\[  \sum_{j=1}^{n-1} (2j-1) + 2(n-1) = n^2 -1 \ . \]

As an example consider the simplest system, i.e., a qubit
corresponding to $n=2$. One obtains \cite{Erwin}
\begin{equation}\label{}
    A^2_{2} = V^2_{2} = \left( \begin{array}{cc} c |\widetilde{z}\>\<\widetilde{z}|   & s |\widetilde{z}\> \\
    -s \<\widetilde{z}| & c
    \end{array} \right) \ ,
\end{equation}
with $|\widetilde{z}\> = e^{i\varphi}$ and $c = \cos\theta$, $s =
\sin\theta$. One has therefore
\begin{equation}\label{}
    \rho_2 = \left( \begin{array}{cc} c^2\lambda_1  + s^2 \lambda_2 & sc\, e^{i\varphi}(\lambda_1 - \lambda_2) \\
    sc\, e^{-i\varphi}(\lambda_1 - \lambda_2) & c^2\lambda_2  + s^2
    \lambda_1 \end{array} \right) \ .
\end{equation}
The above parametrization reproduces the standard Bloch ball:
$\varphi$ and $\vartheta=2\theta$ are nothing but the spherical
angles on the unit Bloch sphere, and $r = \lambda_1 - \lambda_2 \in
[0,1]$ is the radial coordinate. For $\lambda_1 =1$ one obtains the
celebrated Bloch sphere of pure states
\begin{equation}\label{}
    |\psi\>\<\psi| = \left( \begin{array}{cc} c^2 &  e^{i\varphi}\, sc \\
     e^{-i\varphi}\, sc & s^2  \end{array} \right) \ ,
\end{equation}
that is
\begin{equation}\label{}
    |\psi\> = \left( \begin{array}{c} \cos\theta \\
    e^{i\varphi}\sin\theta \end{array} \right) \ ,
\end{equation}
up to an overall phase factor.


\section{Composite $n\ot m$ systems} \label{com}

 Consider now a density operator
for the composite system living in $\mathbb{C}^n \ot \mathbb{C}^m$.
It is clear that we may parameterize it as a density operator living
in $\mathbb{C}^{nm}$. However, this way we loose information about
the particular tensor product structure of the total Hilbert space
$\mathbb{C}^{nm}$. To control the division into subsystems
$\mathbb{C}^{nm} = \mathbb{C}^n \ot \mathbb{C}^m$ let us consider
$\rho$ as an $n \times n$ matrix with $m \times m$ blocks, i.e.
\begin{equation}\label{blocks}
    \rho_{n,m} = \sum_{i,j=1}^n  |i\>\<j|\ot \rho_{ij} \ ,
\end{equation}
with $\rho_{ij}$ being $m \times m$ complex matrices. Our aim is
to provide a suitable parametrization for positive block matrices.
Let $D(\lambda_1,\ldots,\lambda_{nm})$ denote a  diagonal $nm
\times nm$ matrix with $\lambda_i\geq 0$ and $\sum_i \lambda_i=1$.
It is clear that
\begin{equation}\label{}
    \rho_{n,m} = U_{n,m}\cdot D(\lambda_1,\ldots,\lambda_{nm}) \cdot U_{n,m}^\dagger\ ,
\end{equation}
where $U_{n,m} \in SU(nm)$. Any special unitary matrix $U_{n,m}$ may be
written as
\begin{equation}\label{}
    U_{n,m} = e^{X}\ ,
\end{equation}
where $X$ is an $nm\times nm$ anti-hermitian matrix and hence it
may be represented as follows
\begin{equation}  \label{X}
    X = X_1 + X_2 + \ldots + X_n \ ,
\end{equation}
where $X_1$ is anti-hermitian block-diagonal and $X_j$ for $j\geq
2$ are $n\times n$ block anti-hermitian matrices with $m\times m$
blocks defined as follows:
\begin{equation}\label{Xj}
X_j = \left( \begin{array}{c|c|c} \mathbb{I}_{j-1} \ot \mathbb{O}_m & |Z_j\> & 0  \\
\hline - \< Z_j| & \mathbb{O}_m & 0 \\ \hline 0 & 0 &
\mathbb{I}_{n-j} \ot \mathbb{O}_m \end{array} \right) \ ,
\end{equation}
where, instead of $(n-1)$ column vectors $z_j$ we take $(n-1)$
column block vectors
\begin{equation}\label{}
    |Z_j\> = \left( \begin{array}{c} Z_{1,j} \\ \vdots \\ Z_{j-1,j}
    \end{array} \right)\ ,
\end{equation}
with $Z_{i,j}$ being $m \times m$ matrices. Similarly
\[  \<Z_j| = (Z_{1,j}^\dagger,Z_{2,j}^\dagger, \ldots,
Z_{j-1,j}^\dagger)\ . \] Using the parametrization of block
anti-hermitian matrices (\ref{X}) and (\ref{Xj}) we are ready to
define the following parametrization of the unitary group:
\begin{equation}\label{Aj}
    A^j_{n,m} = e^{X_j}\ .
\end{equation}
Hence, we consider unitary matrices from $SU(nm)$
of the following form
\begin{equation}\label{}
U_{n,m} = A^n_{n,m} A^{n-1}_{n,m} \ldots A^2_{n,m}A^1_{n,m}\ ,
\end{equation}
where $A^j_{n,m}$ are unitary block matrices (\ref{Aj}) and
$A^1_{n,m}$ is unitary block diagonal, i.e.,
\begin{equation}\label{}
A^1_{n,m} = \left( \begin{array}{c|c|c|c} U_1 & \mathbb{O}_m & \ldots &  \mathbb{O}_m  \\
\hline \mathbb{O}_m & U_2 & \ldots & \mathbb{O}_m  \\ \hline \vdots & \vdots & \ddots & \vdots \\
\hline \mathbb{O}_m & \mathbb{O}_m & \ldots & U_n
\end{array} \right) \ ,
\end{equation}
where $U_k$ are $m \times m$ unitary matrices. It is clear that
\[  A^1_{n,m} D(\lambda_1,\ldots,\lambda_{nm}) A^{1\, \dagger}_{n,m}  \ ,
\]
is a positive block diagonal matrix. Let us denote it by
$D(\Lambda_1|\ldots|\Lambda_n)$, where $\Lambda_k$ stand for
$m\times m$ diagonal positive blocks
\begin{equation}\label{}
    \Lambda_k = U_k D(\lambda_{km},\ldots,\lambda_{km+m-1}) U_k^\dagger\
    .
\end{equation}
If $m=1$ one has $\Lambda_k = \lambda_k \geq 0$. Moreover, we add
the normalization condition
\begin{equation}\label{TR}
    \mbox{Tr}\, (\Lambda_1 + \ldots + \Lambda_n) = 1 \ .
\end{equation}
One has finally
\begin{equation}\label{}
    \rho_{n,m} = A^n_{n,m} \ldots A^2_{n,m}
    D(\Lambda_1|\ldots|\Lambda_n) A^{2\dagger}_{n,m} \ldots
    A^{n\dagger}_{n,m}\ .
\end{equation}
To apply the above formula one needs the explicit form of the unitary
components $A^j_{n,m}$.
 A straightforward calculation gives (we follow
\cite{Fujii})
\begin{equation}\label{}
    A^j_{n,m} = \left( \begin{array}{c|c|c|c}
    V^j_{n,m} &\mathbb{O}_m &\ldots & \mathbb{O}_m \\ \hline
    \mathbb{O}_m  & \mathbb{I}_m & \ldots & \mathbb{O}_m\\ \hline
    \vdots & \vdots & \ddots & \vdots\\  \hline
   \mathbb{O}_m &\mathbb{O}_m &\ldots  &  \mathbb{I}_m\end{array} \right) \ ,
\end{equation}
where the unit block $\mathbb{I}_m$ appears $n-j$ times. In the
above formula $V^j_{n,m}$ is a $j\times j$ block unitary matrix with
$m\times m$ blocks defined as follows:
\begin{equation}\label{V...}
    V^j_{n,m} = \left( \begin{array}{c|c} \mathbb{I}_{j-1} \ot \mathbb{I}_m -
    |\widetilde{Z}_j\>\Big[ \mathbb{I}_{j-1} \ot (\mathbb{I}_m-C_j)\Big]\<\widetilde{Z}_j|
      & |\widetilde{Z}_j\> S_j \\ \hline
    - S_j\<\widetilde{Z}_j| & C_j
    \end{array} \right) \ ,
\end{equation}
and  $|\widetilde{Z}_j\>$ denotes the normalized block vectors,
that is,
\begin{equation}\label{}
    \widetilde{Z}_{k,j} := \frac{Z_{k,j}}{||Z_j||}\ ,
\end{equation}
where
\begin{equation}\label{}
||Z_j||^2 =    Z_{1,j}^\dagger Z_{1,j} + \ldots + Z_{j-1,j}^\dagger
Z_{j-1,j}\ .
\end{equation}
Moreover, $|\widetilde{Z}_j\>\Big[ \mathbb{I}_{j-1} \ot
(\mathbb{I}_m-C_j)\Big]\<\widetilde{Z}_j| $ stands for the following
$(j-1)\times (j-1)$ block matrix
\begin{equation}\label{}
\sum_{k,l=1}^{j-1} |k\>\<l| \ot \widetilde{Z}_{k,j}^\dagger C_j\,
\widetilde{Z}_{l,j}\    ,
\end{equation}
and
\begin{equation}\label{}
C_j = \cos \Xi_j\ , \ \ \ \  S_j = \sin \Xi_j\ ,
\end{equation}
with
\begin{equation}\label{}
    \Xi_j := ||Z_j||\ .
\end{equation}
It is clear that for $m=1$ we recover the parametrization used in
\cite{Erwin}.


\section{Examples}

\noindent {\it Class 1.}  Let us consider a $2 \ot 2$ system to
illustrate our parametrization for the  well known 2-qubit states.
Taking
\begin{equation}\label{E+++}
    \Lambda_1 = \mathbb{O}_2 \ , \ \ \ \Lambda_2 = \frac 12 (\mathbb{I}_2 - \sigma_z)\ , \
    \ \  S = \sin\alpha\, \mathbb{I}_2\ , \ \ \ \ C= \cos\alpha\,
    \mathbb{I}_2\ , \ \ \ U=\sigma_x \ ,
\end{equation}
one obtains a family of rank-1 projectors
\begin{equation}\label{}
   P(\alpha) = \left( \begin{array}{cc|cc} \sin^2\alpha & 0 & 0 &
    \sin\alpha\cos\alpha \\  0 & 0 & 0 & 0 \\ \hline   0 & 0 & 0 & 0 \\ \sin\alpha\cos\alpha &  0 &
    0 & \cos^2\alpha \end{array} \right) \ ,
\end{equation}
which corresponds to a pure state
$$ \psi_\alpha = \sin\alpha\, |00\> + \cos\alpha\, |11\>\ . $$
Note, that this state is separable if and only if $S=0$ or $C=0$.
For $S=C = \mathbb{I}_2/\sqrt{2}\,$, one obtains a maximally
entangled state. It shows that a nontrivial rotation by $\alpha$
does produce quantum entanglement. As a second example in this class
let us take $S=C = \mathbb{I}_2/\sqrt{2}\,$, $U=\sigma_x$ and
\begin{equation}\label{LL}
    \Lambda_1 = \frac 14 \left( \begin{array}{cc} 1-p & 0 \\ 0 &
    1-p \end{array} \right) \ , \ \ \ \
    \Lambda_2 = \frac 14 \left( \begin{array}{cc} 1-p & 0 \\ 0 &
    1+3p \end{array} \right) \ ,
\end{equation}
with $-1/3 \leq p \leq 1$ to guarantee positivity of the matrices
$\Lambda$. One obtains the following 1-parameter family of 2-qubit
states
\begin{equation}\label{}
   I(p) = \frac 14 \left( \begin{array}{cc|cc} 1+p & 0 & 0 &
    2p \\  0 & 1-p & 0 & 0 \\ \hline   0 & 0 & 1-p & 0 \\ 2p &  0 &
    0 & 1+p \end{array} \right) \ .
\end{equation}
This is the well known family of isotropic states which is known to
be separable if and only if  $p \leq 1/3$. Actually, a point $p=1/3$
is not distinguished by our  parametrization.

 Taking $S=\sin\alpha \,
\mathbb{I}_2$ and $C =\cos\alpha \, \mathbb{I}_2$ one obtains a more
general 2-parameter family
\begin{equation}\label{}
I(p,\alpha) = \frac{1-p}{4}\, \mathbb{I}_2 \ot \mathbb{I}_2 +
pP(\alpha)\ ,
\end{equation}
which is separable if and only if
$$ p\, \leq\, \frac{1}{1 + 2\sin(2\alpha)}\ . $$

The above example may be generalized  as follows. Instead of
(\ref{LL}) let us consider
\begin{equation}\label{LL2}
    \Lambda_1 =  \left( \begin{array}{cc} p_2 & 0 \\ 0 &
    p_4 \end{array} \right) \ , \ \ \ \
    \Lambda_2 =  \left( \begin{array}{cc} p_3 & 0 \\ 0 &
    p_1 \end{array} \right) \ ,
\end{equation}
with $p_k \geq 0$ and $p_1+p_2+p_3+p_4=1$. Taking
$$ S =  \left( \begin{array}{cc} \sin\alpha & 0 \\ 0 &
    \sin\beta \end{array} \right)\ ,\ \ \ \ \
C =  \left( \begin{array}{cc} \cos\alpha & 0 \\ 0 &
    \cos\beta \end{array} \right)\ , \ \ \ \ \ \ \alpha,\beta \in [0,\pi/2]\ , $$
 and $U=\sigma_x$ one obtains the following  family
\begin{equation}\label{}
    \rho({\bf p};\alpha,\beta) =  \left( \begin{array}{cc|cc} p_1 c_\alpha^2+p_2 s_\alpha^2 & 0 & 0 &
    (p_1-p_2)s_\beta c_\beta \\  0 & p_3 c_\beta^2 + p_4 s_\beta^2 & (p_3-p_4)s_\alpha c_\alpha & 0 \\
    \hline   0 & (p_3-p_4)s_\alpha c_\alpha & p_3 s_\beta^2 +p_4 c_\beta^2 & 0 \\ (p_1-p_2)s_\beta c_\beta
     &  0 &
    0 & p_1 s_\alpha^2 +p_2 c_\alpha^2 \end{array} \right) \ ,
\end{equation}
where
$$ s_\alpha = \sin\alpha\ , \ \ \ \ c_\alpha = \cos\alpha\ , $$
and similarly for $s_\beta,c_\beta$. We stress that one has
$\rho({\bf p};\alpha,\beta) \geq 0$ and ${\rm Tr}\, \rho({\bf
p};\alpha,\beta) =1$ for any $\alpha, \beta$ and the arbitrary
probability vector ${\bf p}=(p_1,p_2,p_3,p_4)$ by construction.
Interestingly, the above family belongs to the class of $2 \ot 2$
circulant states considered in \cite{CIRCULANT}. The Peres PPT
criterion \cite{Peres} gives the following separability conditions
\begin{eqnarray}  \label{PPT1}
   p_3 c_\beta^2 + p_4 s_\beta^2 & \geq & |p_1-p_2|s_\beta c_\beta\ ,
   \\ \label{PPT2}
  p_1 c_\alpha^2 + p_2 s_\alpha^2 & \geq & |p_3-p_4|s_\alpha
  c_\alpha\ .
\end{eqnarray}
 For $\alpha =\beta = \pi/4$ the above family reduces to the family of Bell
diagonal states
\begin{equation}\label{}
    \rho({\bf p}) = \frac 12\,  \left( \begin{array}{cc|cc} p_1 + p_2  & 0 & 0 &
    p_1-p_2 \\  0 & p_3  + p_4  & p_3-p_4 & 0 \\
    \hline   0 & p_3-p_4  & p_3  + p_4  & 0 \\ p_1-p_2
     &  0 &  0 & p_1  + p_2  \end{array} \right) \ .
\end{equation}
Moreover, separability conditions (\ref{PPT1})--(\ref{PPT2}) reduce
to $p_k \leq 1/2$ for $k=1,2,3,4$. Note, that even if $\rho({\bf
p})$ is entangled $\rho({\bf p};\alpha,\beta)$ might be separable.
Consider e.g. $p_1=p_2=p_3=1/8$ and $p_4=5/8$, that is,  $\rho({\bf
p})$ is entangled. Now, (\ref{PPT1}) is trivially satisfied  and
(\ref{PPT2}) implies $\sin 2\alpha \leq 1/2$. Hence, $\rho({\bf
p};\alpha,\beta)$ is separable for $\alpha \leq \pi/12$ and
arbitrary $\beta$.

\vspace{.5cm}

 \noindent {\it Class 2.} An arbitrary state of a
$2\ot m$ system corresponds to
\begin{equation}\label{}
    A^2_{2,m} = V^2_{2,m} = \left( \begin{array}{cc} \widetilde{Z} C\, \widetilde{Z}^\dagger   &  \widetilde{Z} S \\
    - S\widetilde{Z}^\dagger & C
    \end{array} \right) \ ,
\end{equation}
with $\widetilde{Z} = U \in U(m)$ and again $C = \cos \Xi_2$, $S =
\sin \Xi_2$. One finds
\begin{eqnarray}\label{}
    \rho_{2,m} &=& \left( \begin{array}{c|c}   U(CU^\dagger  \Lambda_1 UC + S \Lambda_2 S)U^\dagger &
                                            U(S \Lambda_2 C - CU^\dagger \Lambda_1 US) \\ \hline
(C \Lambda_2 S - SU^\dagger \Lambda_1 UC)U^\dagger  &  C \Lambda_2 C
+ SU^\dagger \Lambda_1 US \end{array} \right) \nonumber \\
&=& \left( \begin{array}{c|c} U & \mathbb{O}_m \\ \hline
\mathbb{O}_m & \mathbb{I}_m
\end{array} \right) \left( \begin{array}{c|c}   CU^\dagger  \Lambda_1 UC + S \Lambda_2 S &
                                            S \Lambda_2 C - CU^\dagger \Lambda_1 US \\ \hline
C \Lambda_2 S - SU^\dagger \Lambda_1 UC  &  C \Lambda_2 C +
SU^\dagger \Lambda_1 US \end{array} \right)  \left(
\begin{array}{c|c} U^\dagger & \mathbb{O}_m \\ \hline \mathbb{O}_m &
\mathbb{I}_m
\end{array} \right)\ .
\end{eqnarray}
Note that for $S=0$ or $C=0$ one obtains a class of block-diagonal
matrices
\begin{equation}\label{}
     \left(
\begin{array}{c|c} \Lambda_1 & \mathbb{O}_m \\ \hline \mathbb{O}_m &
\Lambda_2
\end{array} \right)\  \ \ \ \ \mbox{or} \ \ \ \ \    \left(
\begin{array}{c|c} \Lambda_2 & \mathbb{O}_m \\ \hline \mathbb{O}_m &
\Lambda_1 \end{array} \right)\ ,
\end{equation}
respectively. Being block-diagonal these matrices represent
separable $2\ot m$ states. It shows that quantum entanglement arises
only for nontrivial $\Xi_2$ corresponding to $C\neq 0$ and $S \neq
0$.

Note, that for
\begin{enumerate}
\item
$\Lambda_1=\Lambda_2=\Lambda\ , $

\item $[\Lambda,U]=0\ , $

\end{enumerate}
one obtains the following class of $2 \ot m$ states:
\begin{equation}\label{}
    \left(
\begin{array}{c|c} U & \mathbb{O}_m \\ \hline \mathbb{O}_m &
\mathbb{I}_m
\end{array} \right) \left( \begin{array}{c|c}  A & B \\ \hline
 B^\dagger &  A \end{array} \right) \left(
\begin{array}{c|c} U^\dagger & \mathbb{O}_m \\ \hline \mathbb{O}_m &
\mathbb{I}_m
\end{array} \right) \ ,
\end{equation}
 with
\[ A = C  \Lambda  C + S \Lambda S\ , \ \ \ \ B= S\Lambda C - C\Lambda
S\ . \] Now, if $UAU^\dagger = A\,$, then one gets
\begin{equation}\label{}
 \left( \begin{array}{c|c}  A & UB \\ \hline
 (UB)^\dagger &  A \end{array} \right) \ .
\end{equation}
These are  block Toeplitz positive matrices and it is well known
that they are separable \cite{G}. In this way we define huge family
of bipartite separable states. Another class of separable states is
defined by block Hankel positive matrices \cite{G}: taking
$U,\Lambda_1,\Lambda_2$ and $\Xi_2$ satisfying
\begin{enumerate}
\item
$ [U^\dagger \Lambda_1U,\Xi_2]=0 \ , $

\item $[\Lambda_2,\Xi_2]=0\ , $

\end{enumerate}
one obtains the following class of $2 \ot m$ states:
\begin{equation}\label{}
    \left(
\begin{array}{c|c} U & \mathbb{O}_m \\ \hline \mathbb{O}_m &
\mathbb{I}_m
\end{array} \right) \left( \begin{array}{c|c}  A_1 & B' \\ \hline
 B' &  A_2 \end{array} \right) \left(
\begin{array}{c|c} U^\dagger & \mathbb{O}_m \\ \hline \mathbb{O}_m &
\mathbb{I}_m
\end{array} \right) \ ,
\end{equation}
 with
\[ B'= SC(\Lambda_2 - U^\dagger\Lambda_1 U)\ . \]
Now, if $$ UB' = B'U^\dagger\ , $$  then one gets
\begin{equation}\label{}
\left( \begin{array}{c|c}  UA_1U^\dagger & X \\ \hline
 X &  A_2 \end{array} \right) \ ,
\end{equation}
with $X := UB'$. These are  block Hankel positive matrices and hence
separable \cite{G}.

\vspace{.5cm}

\noindent {\it Class 3.} An interesting class of bipartite $n \ot m$
states corresponds to $\Lambda_1 = \ldots
=\Lambda_{n-1}=\mathbb{O}_m$ and $\Lambda_n = \frac 1m
\mathbb{I}_m$. In this case one has
\begin{equation}\label{ADA}
    \rho_{n,m} = \frac 1m A^n_{n,m}\, D(\mathbb{O}_m|\ldots|\mathbb{O}_m|\mathbb{I}_m) \,
    A_{n,m}^{n\dagger}\ .
\end{equation}
This class is parameterized by a positive matrix  $\Xi_n$ and
$(n-1)$ complex matrices
$\widetilde{Z}_{1,n},\ldots,\widetilde{Z}_{n-1,n}$ satisfying
$\sum_{k=1}^{n-1} \widetilde{Z}_{k,n}^\dagger \widetilde{Z}_{k,n} =
\mathbb{I}_m$. Taking into account a polar decomposition
$\widetilde{Z}_{k,n} = P_k U_k$, with positive $P_k$ and unitary
$U_k$, one replaces the above constraint by
\begin{equation}\label{N-sphere}
P_1^2 + \ldots + P_{n-1}^2 = \mathbb{I}_m \ .
\end{equation}
The above equation defines a {\it nonabelian sphere} \cite{A}.
Therefore, to parameterize the class defined by (\ref{ADA}) one may
use $(n-1)$ unitaries $\{U_1,\ldots,U_{n-1}\}$ and $(n-1)$ positive
operators $\{P_1,\ldots,P_{n-2},\Xi_{n}\}$ (since $P_{n-1}$ may be
calculated from (\ref{N-sphere})). Note, that for fixed
$\{P_1,\ldots,P_{n-2},\Xi_n\}$ one obtains a $(n-1)m^2$--dimensional
subspace which may be called a nonabelian $(n-1)$--torus. It is,
therefore, clear that a class (\ref{ADA}) generalizes the subspace
of pure states for $n$-level single system and hence it may be
regarded as a nonabelian generalization of the complex projective
space $\mathbb{C}P^{n-1}$ \cite{Nasza,Karol}. Let us observe that if
$Z_{k,j}$ are normal, that is
\begin{equation}\label{}
    Z_{k,j} Z_{k,j}^\dagger = Z_{k,j}^\dagger Z_{k,j}\ ,
\end{equation}
 then (\ref{ADA}) defines a rank-$m$ projector which generalizes
rank-1 projector (a pure state) for a single $n$-level system. In
particular for $n=2$ one generalizes a 2-dimensional Bloch sphere
(one may call it a nonabelian Bloch sphere):
\begin{equation}\label{}
  \frac 1m  \left( \begin{array}{c|c}   S^2  &
                                               USC  \\ \hline
  CS U^\dagger   &  C^2 \end{array} \right) \ ,
\end{equation}
which is parameterized by two nonabelian angles: unitary $U$ and
positive $\Xi_2$ ($C = \cos \Xi_2$, $S = \sin\Xi_2$). All together
$2m^2$ parameters. Note that for $n=m$ the class (\ref{ADA}) defines
a set of extremal states of the extended quantum theory proposed
recently by \.Zyczkowski \cite{Zyczkowski}.


\section{Conclusions}

We proposed a parametrization of density matrices of composed $n
\ot m$ quantum system. For $m=1$ this parametrization reduces to
the one used recently in \cite{Erwin}. Note, that it may be
generalized for multipartite systems living in $\mathbb{C}^{n_1}
\ot \ldots \ot \mathbb{C}^{n_N}$. Indeed, instead of dealing with
$n\times n$ block matrices with $m\times m$ blocks, in the
multipartite case one has to consider $n_1 \times n_1$ block
matrices with blocks being $n_2 \times n_2$ block matrices with
blocks being $n_3\times n_3$ block matrices and so on. Although
the strategy seems to be simple the technical part of the story is
quite involved. It is anticipated that the presented
parametrization will find interesting applications in the study of
quantum properties of many partite systems.

\vspace{.5cm}

\noindent {\it Acknowledgement} This work is based upon research
supported by the South African Research Chairs Initiative of the
Department of Science and Technology and National Research
Foundation. One of us (DC) was partially supported by the Polish
Ministry of Science and Higher Education Grant No 3004/B/H03/2007/33
and by the Polish Research Network  {\it Laboratory of Physical
Foundations of Information Processing}. DC thanks Francesco
Petruccione for the warm hospitality in Durban.


\begin{thebibliography}{1} \bibliographystyle{plain}

\bibitem{BP} H.-P. Breuer and F. Petruccione, {\it The Theory of Open
Quantum Systems}, Oxford University Press, Oxford, 2007.

\bibitem{Alicki} R. Alicki, K. Lendi, {\it Quantum Dynamical Semigroups and
Application, Lecture Notes in Physics},  Vol. {\bf 286},
Springer-Verlag, Berlin, 1987.

\bibitem{QIT}  M. A. Nielsen and I. L. Chuang, {\it Quantum computation
and quantum information}, Cambridge University Press, Cambridge,
2000.

\bibitem{Nasza} D. Chru\'sci\'nski and A. Jamio{\l}kowski, {\it Geometric Phases in Classical and Quantum Mechanics},
Birkh\"auser Boston, 2004.

\bibitem{Karol} I. Bengtsson and K. \.Zyczkowski, {\it Geometry of Quantum States}, Cambridge University Press,  2006.



\bibitem{Hioe} F. T. Hioe and J. H. Eberly, Phys. Rev. Lett. {\bf 47} (1981) 838.


\bibitem{Byrd} M. S. Byrd and N. Khaneja, Phys. Rev. A {\bf 68} (2003)
062322.

\bibitem{Kimura} G. Kimura, Phys. Lett. A {\bf 314} (2003) 339.

\bibitem{Kimura-Kossak} G. Kimura and A. Kossakowski, Open Sys. Information Dyn. {\bf 12} (2005) 207.



\bibitem{Tilma1} T. Tilma and E.C.G. Sudarshan, J. Phys. A: Math. Gen. {\bf 35} (2002)
10467; J. Geom. Phys. {\bf 52} (2004) 263.


\bibitem{Tilma3} T. Tilma, M.S. Byrd and E.C.G. Sudarshan, J. Phys. A: Math. Gen. {\bf 35} (2002) 10445


\bibitem{Dita0} P. Dita, J. Phys. A: Math. Gen. {\bf 15} (1982) 3465; {\bf 36} (2003)
2781; {\bf 38} (2005) 2657.




\bibitem{BK} R. A. Bertlmann and P. Krammer, J. Phys. A: Math. Theor. {\bf 41} (2008)
235303.

\bibitem{Erwin} E. Br\"uning and F. Petruccione, Open Sys. Information
Dyn. {\bf 15} (2008) 109-121.

\bibitem{J} C. Jarlskog, J. Math. Phys. {\bf 46} (2005) 103508;  J. Math. Phys. {\bf 47} (2006)
013507.

\bibitem{Fujii} K. Fujii, K. Funahashi and T. Kobayashi, Int. J. Geom. Meth. Mod. Phys. {\bf 3} (2006)
269.

\bibitem{SPPT} D. Chru\'sci\'nski, J. Jurkowski and A. Kossakowski, Phys. Rev. A {\bf 77} (2008)
022113.






\bibitem{CIRCULANT} D. Chru\'sci\'nski and A. Kossakowski,
 Phys. Rev. A {\bf 76},  032308 (2007).

\bibitem{Peres} A. Peres, Phys. Rev. Lett. {\bf 77}, 1413 (1996).


\bibitem{G} L. Gurvits and H. Barnum,  Phys. Rev. A {\bf 68} (2003)
042312.


\bibitem{A} W. Arveson, {\it The probability of entanglement},
arXiv: 0712.4163.

\bibitem{Zyczkowski}  K. \.Zyczkowski, {\it An extended, quartic quantum theory}, arXiv: 0804.1247.

\end{thebibliography}
\end{document}